\documentstyle[12pt]{article}

\catcode`\@=11
\long\def\@makefntext#1{ 
\protect\noindent \hbox to 3.2pt {\hskip-.9pt
$^{{\ninerm\@thefnmark}}$\hfil}#1\hfill} 

\def\thefootnote{\fnsymbol{footnote}}
 \def\@makefnmark{\hbox to 0pt{$^{\@thefnmark}$\hss}}  

\def\ps@myheadings{\let\@mkboth\@gobbletwo
\def\@oddhead{\hbox{} 
\rightmark\hfil\ninerm\thepage}
\def\@oddfoot{}\def\@evenhead{\ninerm\thepage\hfil 
\leftmark\hbox{}}\def\@evenfoot{}
\def\sectionmark##1{}\def\subsectionmark##1{}}

\begin{document}

\newcommand{\symbolfootnote}{\renewcommand{\thefootnote}
	{\fnsymbol{footnote}}}
\renewcommand{\thefootnote}{\fnsymbol{footnote}}
\newcommand{\alphfootnote}
	{\setcounter{footnote}{0}
	 \renewcommand{\thefootnote}{\sevenrm\alph{footnote}}}

\newcounter{sectionc}\newcounter{subsectionc}\newcounter{subsubsectionc}
\renewcommand{\section}[1] {\vspace{0.6cm}\addtocounter{sectionc}{1}
\setcounter{subsectionc}{0}\setcounter{subsubsectionc}{0}\noindent
	{\bf\thesectionc. #1}\par\vspace{0.4cm}}
\renewcommand{\subsection}[1] {\vspace{0.6cm}\addtocounter{subsectionc}{1}
	\setcounter{subsubsectionc}{0}\noindent
	{\it\thesectionc.\thesubsectionc. #1}\par\vspace{0.4cm}}
\renewcommand{\subsubsection}[1]{\vspace{0.6cm}\addtocounter{subsubsectionc}{1}
	\noindent {\rm\thesectionc.\thesubsectionc.\thesubsubsectionc.
	#1}\par\vspace{0.4cm}}
\newcommand{\nonumsection}[1] {\vspace{0.6cm}\noindent{\bf #1}
	\par\vspace{0.4cm}}

\newcounter{appendixc}
\newcounter{subappendixc}[appendixc]
\newcounter{subsubappendixc}[subappendixc]
\renewcommand{\thesubappendixc}{\Alph{appendixc}.\arabic{subappendixc}}
\renewcommand{\thesubsubappendixc}
	{\Alph{appendixc}.\arabic{subappendixc}.\arabic{subsubappendixc}}

\renewcommand{\appendix}[1] {\vspace{0.6cm}
        \refstepcounter{appendixc}
        \setcounter{figure}{0}
        \setcounter{table}{0}
        \setcounter{equation}{0}
        \renewcommand{\thefigure}{\Alph{appendixc}.\arabic{figure}}
        \renewcommand{\thetable}{\Alph{appendixc}.\arabic{table}}
        \renewcommand{\theappendixc}{\Alph{appendixc}}
        \renewcommand{\theequation}{\Alph{appendixc}.\arabic{equation}}
        \noindent{\bf Appendix \theappendixc #1}\par\vspace{0.4cm}}
\newcommand{\subappendix}[1] {\vspace{0.6cm}
        \refstepcounter{subappendixc}
        \noindent{\bf Appendix \thesubappendixc. #1}\par\vspace{0.4cm}}
\newcommand{\subsubappendix}[1] {\vspace{0.6cm}
        \refstepcounter{subsubappendixc}
        \noindent{\it Appendix \thesubsubappendixc. #1}
	\par\vspace{0.4cm}}

\def\abstracts#1{{
	\centering{\begin{minipage}{30pc}\tenrm\baselineskip=12pt\noindent
	\centerline{\tenrm ABSTRACT}\vspace{0.3cm}
	\parindent=0pt #1
	\end{minipage} }\par}}

\newcommand{\bibit}{\it}
\newcommand{\bibbf}{\bf}
\renewenvironment{thebibliography}[1]
	{\begin{list}{\arabic{enumi}.}
	{\usecounter{enumi}\setlength{\parsep}{0pt}
\setlength{\leftmargin 1.25cm}{\rightmargin 0pt}
	 \setlength{\itemsep}{0pt} \settowidth
	{\labelwidth}{#1.}\sloppy}}{\end{list}}

\topsep=0in\parsep=0in\itemsep=0in
\parindent=1.5pc

\newcounter{itemlistc}
\newcounter{romanlistc}
\newcounter{alphlistc}
\newcounter{arabiclistc}
\newenvironment{itemlist}
    	{\setcounter{itemlistc}{0}
	 \begin{list}{$\bullet$}
	{\usecounter{itemlistc}
	 \setlength{\parsep}{0pt}
	 \setlength{\itemsep}{0pt}}}{\end{list}}

\newenvironment{romanlist}
	{\setcounter{romanlistc}{0}
	 \begin{list}{$($\roman{romanlistc}$)$}
	{\usecounter{romanlistc}
	 \setlength{\parsep}{0pt}
	 \setlength{\itemsep}{0pt}}}{\end{list}}

\newenvironment{alphlist}
	{\setcounter{alphlistc}{0}
	 \begin{list}{$($\alph{alphlistc}$)$}
	{\usecounter{alphlistc}
	 \setlength{\parsep}{0pt}
	 \setlength{\itemsep}{0pt}}}{\end{list}}

\newenvironment{arabiclist}
	{\setcounter{arabiclistc}{0}
	 \begin{list}{\arabic{arabiclistc}}
	{\usecounter{arabiclistc}
	 \setlength{\parsep}{0pt}
	 \setlength{\itemsep}{0pt}}}{\end{list}}

\newcommand{\fcaption}[1]{
        \refstepcounter{figure}
        \setbox\@tempboxa = \hbox{\tenrm Fig.~\thefigure. #1}
        \ifdim \wd\@tempboxa > 6in
           {\begin{center}
        \parbox{6in}{\tenrm\baselineskip=12pt Fig.~\thefigure. #1 }
            \end{center}}
        \else
             {\begin{center}
             {\tenrm Fig.~\thefigure. #1}
              \end{center}}
        \fi}

\newcommand{\tcaption}[1]{
        \refstepcounter{table}
        \setbox\@tempboxa = \hbox{\tenrm Table~\thetable. #1}
        \ifdim \wd\@tempboxa > 6in
           {\begin{center}
        \parbox{6in}{\tenrm\baselineskip=12pt Table~\thetable. #1 }
            \end{center}}
        \else
             {\begin{center}
             {\tenrm Table~\thetable. #1}
              \end{center}}
        \fi}

\def\@citex[#1]#2{\if@filesw\immediate\write\@auxout
	{\string\citation{#2}}\fi
\def\@citea{}\@cite{\@for\@citeb:=#2\do
	{\@citea\def\@citea{,}\@ifundefined
	{b@\@citeb}{{\bf ?}\@warning
	{Citation `\@citeb' on page \thepage \space undefined}}
	{\csname b@\@citeb\endcsname}}}{#1}}

\newif\if@cghi
\def\cite{\@cghitrue\@ifnextchar [{\@tempswatrue
	\@citex}{\@tempswafalse\@citex[]}}
\def\citelow{\@cghifalse\@ifnextchar [{\@tempswatrue
	\@citex}{\@tempswafalse\@citex[]}}
\def\@cite#1#2{{$\null^{#1}$\if@tempswa\typeout
	{IJCGA warning: optional citation argument
	ignored: `#2'} \fi}}
\newcommand{\citeup}{\cite}

\def\fnm#1{$^{\mbox{\scriptsize #1}}$}
\def\fnt#1#2{\footnotetext{\kern-.3em
	{$^{\mbox{\sevenrm #1}}$}{#2}}}

\font\twelvebf=cmbx10 scaled\magstep 1
\font\twelverm=cmr10 scaled\magstep 1
\font\twelveit=cmti10 scaled\magstep 1
\font\elevenbf=cmbx10 scaled 1100
\font\elevenrm=cmr10 scaled\magstephalf
\font\elevenit=cmti10 scaled\magstephalf
\font\tenbf=cmbx10
\font\tenrm=cmr10
\font\tenit=cmti10
\font\ninebf=cmbx9
\font\ninerm=cmr9
\font\nineit=cmti9
\font\eightbf=cmbx8
\font\eightrm=cmr8
\font\eightit=cmti8

\def\BR{{\bf R}}
\def\BC{{\bf C}}
\begin{flushright}
TIFR/TH/94-17\\
hep-th/9405186\\
May, 1994~~~~~~
\end{flushright}
\centerline{\tenbf ON THE GEOMETRY OF $W_n$ GRAVITY\footnote{Talk
presented at the International Colloquium on Modern Quantum Field
Theory II, Tata Institute of Fundamental Research, Bombay during Jan.
5-11,1994.}}
\baselineskip=16pt
\vspace{0.8cm}
\centerline{\tenrm SURESH GOVINDARAJAN\footnote{E-mail:
suresh@theory.tifr.res.in}}
\baselineskip=13pt
\centerline{\tenit Theoretical Physics Group,
Tata Institute of Fundamental Research}
\baselineskip=12pt
\centerline{\tenit Bombay, 400 005, India}
\vspace{0.9cm}
\abstracts{We report work done with T. Jayaraman in this
talk\cite{proposal}.  In a recent paper, Hitchin introduced
generalisations of the Teichmuller space of Riemann surfaces. We
relate these spaces to the Teichmuller spaces of W-gravity. We show
how this provides a covariant description of W-gravity and naturally
leads to a Polyakov path integral prescription for W-strings.}
\baselineskip=14pt
\section{Introduction}
In an interesting paper\cite{hitchin}, Hitchin obtained a class of
Teichmuller spaces as the space of solutions of {\it self-duality}
equations. These spaces have the following features\\

\noindent
(i) They are natural generalisations of the Teichmuller space
of compact Riemann surfaces with genus $g>1$ (which we shall call
${\cal T}_2$).\\
\noindent
(ii) ${\cal T}_2$ embeds in these generalised Teichmuller
spaces.\\
\noindent
(iii) These spaces have the dimension one would expect for
Teichmuller spaces for W-gravity (as given by the zero modes of the
antighosts). For example, for the case of $WA_{(n-1)}$-gravity, the
ghost system consists of $(b^{(j)},c^{(1-j)})$ systems for
$j=2,\ldots,n$, where the superscript refers to the spin of the field.
The number of antighost zero modes (for genus $g>1$) is given by
$$
\sum_{j=2}^n (\#~of~ b^{(j)}~{\rm zero ~modes}) = (2g-2)\sum_{j=2}^n
(2j-1) =(2g-2)~{\rm dim}[ SL(n,\BR)]
$$

In this talk, I will demonstrate the relation of the Teichmuller
spaces of Hitchin to W-gravity. As we shall see this will enable us to
give a gauge independent description of W-gravity using
generalisations of vielbeins and spin-connection of usual gravity.

Before that let me discuss some open questions in W-gravity. First,
from the viewpoint of string theory, one would like to know if there
are any non-trivial examples of W-strings.  The known ones are
generalisations of ``Liouville theory''which suggest that the c=1
barrier exist even in the case of W-strings\cite{wstring}. Another
question which arises is whether there is any geometric structure
underlying W-symmetry. This geometric structure could be of the form
of extra data on the Riemann surface or some immersion in a higher
dimensional manifold. Also, we would like to have a gauge independent
description of W-symmetry. An important feature of W-symmetry is its
non-linear nature. Can we provide a setting where these non-linear
transformations become linear? Can we introduce a $w$-coordinate like
the Grassmann coordinates in supergravity? Is there a Polyakov
path-integral prescription for W-strings? In this talk I will address
some of these issues and hope to address them all in the future. Of
course, I must point out that some of these questions have already
been addressed in some form or the other in existing
literature. However, no one has provided a complete answer to all
these questions.

The plan of the talk is as follows: First, I will briefly discuss
various definitions of the Teichmuller space of Riemann surfaces.
Then, I will introduce {\it self-duality equations} and illustrate it
using the bosonic string. This will provide another definition for
Teichmuller space.  We will then generalise to the case of arbitrary
W-gravity but will use $W_3=WA_2$ to give more details.  Finally, I
will discuss how W-diffeomorphisms arise in this formulation.

\section{Various definitions of Teichmuller space}
Consider a compact Riemann surface $\Sigma$ of genus $g>1$.
Teichmuller space can be defined as follows
\begin{eqnarray}
{\cal T}_2 (\Sigma) &=&
{{\{ {\rm space~of~all~metrics~on~} \Sigma
\}}\over{{\{\rm diff}\}_0 \times \{{\rm Weyl}\}}}\\
&=&{{\{ {\rm space~of~all~metrics~with~constant~negative~curvature~on~} \Sigma
\}}\over{{\{\rm diff}\}_0}}\\
&=&{\rm a~component~of~the~space~Hom}(\pi_1(\Sigma);PSL(2,\BR))/PSL(2,\BR)~,
\end{eqnarray}
where \{diff\}$_0$ refers to diffeomorphisms connected to the
identity.
The definition given in the third line follows from the uniformisation
theorem. We know that $\Sigma$ can be described by quotient
of the upper half plane (with the standard Poincare metric) by a
discrete subgroup (Fuchsian) of $PSL(2,R)$ which is isomorphic to
$\pi_1(\Sigma)$. This provides a homomorphism of $\pi_1(\Sigma)$ to
$PSL(2,\BR)$ which is defined upto conjugation in $PSL(2,\BR)$ (since two
Fuchsian subgroups give the same surface if they belong to the same
conjugacy class). However the space Hom$(\pi_1(\Sigma);PSL(2,\BR))/PSL(2,\BR)$
has many disconnected components. The component of interest is
specified by requiring that the first Chern class of the $U(1)$ part
of the flat $PSL(2,\BR)$ bundle (associated with every homomorphism) is
equal to $(2g-2)$.

\section{Self-duality equations}
The self-duality equations (SDE) are
\begin{equation}
F_A + [\Phi,\Phi^\dagger]=0\quad,\label{sdea}
\end{equation}
where $A$ is a unitary connection on a holomorphic vector bundle $V$,
$F_A$ is its field strength. $\Phi$ is a holomorphic section of
End$(V)\otimes K$ ($K$ is the canonical line bundle on $\Sigma$).
$(V,\Phi)$ form a {\it Higgs bundle}. If a certain stability condition
is satisfied and $c_1(V)=0$, then the connection $A$ (which is
compatible with the holomorphic structure) satisfying (\ref{sdea}) is
unique\cite{selfdual,hitchin}. The holomorphicity condition on $\Phi$
is
\begin{equation}
\bar{\partial}_A \Phi =0\quad. \label{sdeb}
\end{equation}
Equations (\ref{sdea}) and (\ref{sdeb}) imply that ${\cal A}\equiv A +
\Phi + \Phi^\dagger$ is generically a flat $GL(n,\BC)$
connection\footnote{However, in the examples we will discuss here, one
obtains a flat $SL(n,\BC)$ connection}. We shall now illustrate how
SDE's occur in the bosonic as well as W-strings.

\subsection{Bosonic String}

Choose $V=K^{-{1\over2}}\oplus K^{1\over2}$ and
\begin{equation}\label{ehiggsa}
\Phi=\pmatrix{0&h\cr0&0}~dz\quad,
\end{equation}
where the metric on $\Sigma$ is given by $g=h^2~dzd{\bar z}$.
Here $A$ is a $SU(2)$ connection and ${\cal A}=A+\Phi+\Phi^\dagger$ is
a flat $PSL(2,\BC)$ connection. The self-duality equations (\ref{sdea})
imply
\begin{equation}
F_A = \pmatrix{{1\over2}F^0&0\cr0&-{1\over2}F^0}=\pmatrix{1&0\cr0&-1}
h^2dz\wedge d{\bar z}\quad.
\end{equation}
{}From the above equation, we can see that the $SU(2)$ connection is
reducible to a $U(1)$ connection which we will identify with the {\it
spin-connection} on $\Sigma$. Then the condition on $F^0$ is nothing
but the constant curvature condition on the metric.
Hitchin\cite{hitchin} has shown that the connection ${\cal A}$ has
holonomy contained in a real form of $PSL(2,\BC)$ which is
$PSL(2,\BR)$.  In all the examples considered here, the holonomy will
be contained in a real form. Hence, we shall treat ${\cal A}$ as a
flat $PSL(2,\BR)$ connection.

Deform $\Phi$ to
\begin{equation}
\Phi=\pmatrix{0&h\cr{a\over h}&0}~dz\quad,\label{ehiggsb}
\end{equation}
where $a\in{\rm Hom}(K^{-{1\over2}};K^{1\over2})\otimes K = K^2$ is a
holomorphic (follows from equation (\ref{sdeb})). Again, it can be
shown that the self-duality equations imply the constant negative
curvature condition. This leads us to the following conclusion. {\it
The space of solutions of SDE connected to $\Phi=\pmatrix{0&h\cr0&0}$
is the same as the space of constant negative curvature metrics.} As
already discussed this space is ${\cal T}_2$. So we now have an
alternate definition for ${\cal T}_2$.

We shall now make a direct connection with 2d gravity. We have already
identified the $U(1)$ connection $A$ with the spin-connection on
$\Sigma$. We make another identification\cite{tftreview}
\begin{equation}
\Phi + \Phi^\dagger = {\rm vielbeins}~e^+,~e^-~on~\Sigma\quad.
\end{equation}
This implies that $(A,e^+,e^-)$ form a flat $PSL(2,\BR)$ connection.
This has been observed earlier in the context of topological
gravity\cite{topgrav} and later by H. Verlinde\cite{verlinde}. With
this identification, one can see that the holomorphicity condition on
the Higgs field $\Phi$ correspond to the torsion constraints imposed
on the vielbein (in the conformal gauge). The metric $g_2$ is given by
the quadratic $SO(2)$ invariant -- $g_2 = tr(E^2)$, where
$E\equiv\Phi+\Phi^\dagger$.

\subsection{W String}

The generalisation to the W-string case is now easy. We shall restrict
our discussion to the $WA_(n-1)$ string for simplicity.  Following
Hitchin\cite{hitchin}, we consider the vector bundle $V_n$ given by
\begin{equation}
V_n\equiv S^{n-1}(K^{-{1\over2}}\oplus K^{1\over2})
=K^{-{{n-1}\over2}}\oplus K^{-{{n-3}\over2}}\oplus \ldots \oplus
K^{{{n-1}\over2}}
\end{equation}
and choose the Higgs field
\begin{equation}\label{ehiggsc}
\Phi_n=\pmatrix{   0   &   h    &   0  &\ldots& 0\cr
                {a_1 \over h}  & \ddots &\ddots&\ddots&\vdots \cr
	      {a_{2}\over h^2}  & \ddots &\ddots&  h  & 0\cr
	      \vdots & \ddots & {a_1\over h}  &  0  & h \cr
{a_{n-1}\over h^{(n-1)}}& \ldots & {a_2\over h^2}  & {a_1\over h} & 0}~dz\quad,
\end{equation}
where $a_i\in K^{i+1}$. Again ${\cal A}\equiv A + \Phi + \Phi^\dagger$
is a flat $PSL(n,\BC)$ connection provided the self-duality equations
are satisfied. However, the flat connection ${\cal A}$ has its
holonomy contained in a split real form of $PSL(n,\BC)$ which is
isomorphic to $PSL(n,\BR)$.

The bosonic string example suggests that we make the following
identifications
\begin{center}
{\it
{}~~~~~~~~A ---  generalised spin-connection for W-gravity.\\
$\Phi+\Phi^\dagger$ --- generalised vielbein for W-gravity.}
\end{center}

Does this make sense? First, consider the space
\begin{center}
${\cal T}_n\equiv$
\{a component of the space of solutions of SDE for
$(V_n,\Phi_n)$\}\footnote{This component corresponds to the {\it
Teichmuller} component of Hitchin\cite{hitchin}.}.
\end{center}
This component has dimension $(2g-2)$dim$[SL(n,R)]$, which as we have
seen earlier is the dimension we expect for the dimension for the
Teichmuller space for $WA_n$ gravity. Further, local deformations are
given by quadratic, cubic, \ldots differentials. Also, the presence of
higher order invariants lets us define higher order symmetric tensors.
For example, for $n=3$, the quadratic $SO(3)$ invariant gives the
metric ($g_2\equiv tr(E^2$)) and the cubic one gives a symmetric third
rank tensor ($g_3\equiv tr(E^3$)).  So this seems to suggest that the
identifications are sensible. In the So this seems to suggest that
these identifications are sensible. In the latter part of the talk, we
shall demonstrate how the ``usual'' W-diffeomorphisms are recovered in
the conformal gauge. The self-duality equations (\ref{sdea}) and
(\ref{sdeb}) can now be interpreted as generalisations of the constant
curvature condition and torsion constraints respectively.

In the bosonic case, the constant curvature condition corresponded to
gauge fixing the Weyl degree of freedom. So restoring the Weyl degree
corresponds to relaxing the constant curvature condition. In a similar
fashion, in the general case, we can restore the the W-Weyl degrees of
freedom by relaxing the generalised constant curvature condition
(which is given by equation (\ref{sdea})). Further, the W-Weyl
transformations can be obtained using a method due to Howe\cite{howe}.
Here Weyl transformations are seen as transformations which keep the
torsion constraints invariant.

We shall discuss the $W_3$ case in more detail now. The $PSL(3,\BR)$
connection can be parametrised as follows
\begin{equation}
{\cal A} =  \pmatrix{ \omega - {e^{+-}\over \sqrt{3}} &
(\omega^- + e^-) & \sqrt{2}e^{--}\cr
(\omega^+ - e^+) & {{2 e^{+-}} \over \sqrt{3}}&
(-\omega^- + e^-) \cr
\sqrt{2}e^{++}&(-\omega^+ - e^+) & - \omega - {e^{+-}\over
\sqrt{3}} }
\end{equation}
where $\omega, \omega^\pm$ form the $SO(3)$ connection and $e^*$ are
the generalised vielbein. We would like to make the following
observations.
\begin{enumerate}
\item[(i)] Here, $(e^\pm,\omega)$ form a $PSL(2,\BR)$ conn\-ec\-tion which is
em\-bed\-ded in $PSL(3,\BR)$.
\item[(ii)] $e^{++},e^{--},e^{+-}$ are the new vielbein. They have
been labelled by their $U(1)$ charges (w.r.t. $\omega$). Schoutens
{\it et al.} introduced W-vielbein which correspond to $e^{++}$ and
$e^{--}$ (but not for $e^{+-}$) in their covariant construction of an
action for scalar fields coupled to $WA_2$ gravity\cite{schoutens}.
\item[(iii)] The geometry is not Riemannian anymore in the sense that
the torsion constraints are not sufficient to determine the connection
in terms of the vielbein. However, we
can always choose a gauge where we trade one of the gauge symmetries
to determine the connection in terms of the vielbein.
\item[(iv)] Conformal gauge corresponds to choosing
$$ e^+_z=e^-_{\bar z}=h\quad,\quad e^{++}_z=e^{--}_{\bar z}=0\quad, $$
which fixes the $SO(3)$ gauge freedom. Further, these gauge choices
are algebraic and hence their corresponding ghosts can be ignored
(since they would be non-interacting). Next, choose the following
gauge choices
$$e^+_{\bar z}=e^-_z=e^{++}_{\bar z}=e^{--}_z=0\quad,$$
whose corresponding ghosts (anti-ghosts) are of spin $-1,-2$ $(2,3)$
as expected for $W_3$ gravity. Further, the residual transformations
which preserve this gauge choice correspond to holomorphic
(anti - holomorphic) transformations $\epsilon^+,\epsilon^{++}$
($\epsilon^-,\epsilon^{--}$).
\end{enumerate}

\section{W-diffeomorphisms}
The work of Gerasimov {\it et al.}\cite{GLM} and subsequently that of
Bilal {\it et al.}\cite{BFK} has provided a geometric picture of
W-diffeomorphisms in the conformal gauge. They have shown that
W-diffeomorphisms correspond to deformations of certain flag
manifolds. In their construction, the action of $WA_{(n-1)}$
diffeomorphisms on vector space $V=S^{n-1}(K^{-{1\over2}}\oplus
K^{1\over2})$ was demonstrated. This is precisely the space on which
the Higgs field acts. We shall show that this is not a coincidence and
show that provided a certain constraint is obeyed, the $SL(n,\BR)$
gauge transformations are equivalent to W-diffeomorphisms. Other
related works are \cite{others}. The feature which is different in our
formulation here is that one does not need to invoke matter fields in
order to describe W-diffeomorphisms. The role of projective
connections is played by combinations of the generalised spin
connections.

\subsection{The bosonic case}

As usual, the bosonic ($PSL(2,\BR)$) case shows us the way. Choose
 the Higgs field $\Phi$ as in eqn. (\ref{ehiggsb}). Consider the fields
$(\tilde\psi_1,\tilde\psi_2)\in (K^{-1\over2},K^{1\over2})$ subject to
the conditions
\begin{eqnarray}
 ({\bar\partial}_\omega+ \Phi^\dagger )\pmatrix{\tilde\psi_1\cr\tilde\psi_2}=0
\label{econda}\quad,\\
 ({\partial}_\omega+ \Phi)\pmatrix{\tilde\psi_1\cr\tilde\psi_2}=0
\label{econdb}\quad.
\end{eqnarray}
Eqn. (\ref{econda}) implies that the fields $\psi_i$ are holomorphic
while the second eqn. (\ref{econdb}) is a constraint on $\psi_i$.
Interestingly, the self-duality equations (\ref{sdea}) and
(\ref{sdeb}) imply the consistency of the two
conditions we have just imposed. The holomorphicity condition
(\ref{sdeb}) on $\Phi$ implies that
\begin{equation}\label{econna}
\omega_z = -h^{-1}\partial_z~h\quad,\quad \omega_{\bar
z}=h^{-1}\partial_{\bar z}~h\quad,
\end{equation}
and the condition that $a$ is holomorphic ($\partial_{\bar z}~a=0$).
We shall now rescale the fields $\psi_i$ as follows
\begin{equation}\label{escala}
\pmatrix{\tilde\psi_1\cr\tilde\psi_2} = \pmatrix{h^{1\over2}\psi_1\cr
h^{-{1\over2}}\psi_2} \quad.
\end{equation}
$(\psi_1,\psi_2)$ correspond to {\it primary fields} in the
CFT sense, i.e., they have Einstein indices. This rescaling now
enables us to make contact with diffeomorphisms in CFT. Further, the
Christoffel connection are given by $\Gamma_{zz}^{~~z}=2\omega_z$ and
$\Gamma_{{\bar z}{\bar z}}^{~~{\bar z}}=2\omega_{\bar z}$. A simple
calculation shows that conditions (\ref{econda}) and (\ref{econdb})
translates to the following conditions
\begin{eqnarray}
(\partial_{\bar z} +{\bar\mu}\partial_z - {1\over2}\partial_z
{\bar\mu})\psi_1 =0\label{econdaa}\quad,\\
(\partial_z^2 -u)\psi_1 =0 \label{econdba}\quad,
\end{eqnarray}
where ${\bar\mu}\equiv {{\bar a}\over{h^2}}$ is the Beltrami
differential and $u\equiv {1\over2}\partial_z\Gamma_{zz}^{~~z}
+{1\over4}(\Gamma_{zz}^{~~z})^2 -a$.  It is a simple exercise to check
that $u$ transforms like the Schwarzian. A similar observation was
made by Sonoda\cite{sonoda} who pointed out that such a term could be
added to the energy momentum tensor to make it transform like a
$(2,0)$ tensor. Hence, $u$ behaves like a projective
connection. Eqn. (\ref{econdaa}) implies that $\psi_1$ transforms as
$$
\delta~\psi_1=\xi^z\partial_z \psi_1 -{1\over2}(\partial_z \xi^z)\psi_1\quad,
$$
which is the standard transformation of a $(-{1\over2},0)$ tensor in CFT.
What we have done here is similar to what \cite{GLM,BFK} have done.
However, there are some differences. The flatness condition that has
been considered in \cite{GLM,BFK} is different from the flatness
condition implied by the self-duality equations. The $PSL(2,\BR)$
gauge field is a completely geometric object with no relation apriori
to matter fields. However, certain special combinations of the
spin-connection transforms like the Schwarzian. This combination can
be related to the stress-tensor via Ward identities
considered by Verlinde\cite{verlinde,GLM,BFK}. It can also be  seen that
$W-$transformations have a presentation here without directly
involving ``matter fields''.

The compatibility of conditions (\ref{econdaa}) and (\ref{econdba})
implies that
\begin{equation}\label{econsa}
[\partial_z + {\bar \mu}\partial_z + 2 (\partial_z {\bar\mu})]~u =
{1\over2} \partial_z^3{\bar \mu}\quad,
\end{equation}
which is equivalent to the standard OPE for the stress-tensor
(following a procedure outlined in \cite{GLM}.)
\begin{equation}\label{eopea}
T(z)~T(w) \sim {{c/2}\over {(z-w)^4}} + {{2T(w)}\over {(z-w)^2}} +
{{\partial_wT(w)}\over {(z-w)}} +\ldots\quad,
\end{equation}
provided we identify
$ u \longrightarrow {6\over c} \langle T \rangle.$
We thus recover the residual diffeomorphisms which preserve conformal gauge.

\subsection{The $W_3$ case}

We shall now repeat the exercise of the previous subsection for the
$WA_2$ case. Consider the multiplet
$(\tilde\psi_1,\tilde\psi_2,\tilde\psi_3)\in (K^{-1},K^0,K)$. The
Higgs field is given by (\ref{ehiggsc}) to be
$$
\Phi=\pmatrix{0&h&0\cr {a\over h}&0&h\cr {b\over {h^2}}&{a\over h} &0}dz
+ {\cal O}(a^2,ab,b^2)\quad.
$$
The holomorphicity condition (\ref{sdeb}) implies that
\begin{equation}
\omega^+_\mu = \omega^-_\mu =0 \quad,\quad \omega_z =-h^{-1}\partial_z
h\quad,\quad \omega_z =h^{-1}\partial_{\bar z} h\quad,
\end{equation}
and that $(a,b)$ are holomorphic ($\partial_{\bar z}a=\partial_{\bar
  z}b=0$).  The above solution is valid to ${\cal O}(a^2,ab,b^2)$.
Impose the following conditions on $(\tilde\psi_1,\tilde\psi_2,\tilde\psi_3)$.
\begin{eqnarray}
 ({\bar\partial}_\omega + \Phi^\dagger
)\pmatrix{\tilde\psi_1\cr\tilde\psi_2\cr\tilde\psi^3}=0
\label{econdc}\quad,\\
 (\partial_\omega+ \Phi)\pmatrix{\tilde\psi_1\cr\tilde\psi_2\cr\tilde\psi_3}=0
\label{econdd}\quad.
\end{eqnarray}
Eqn. (\ref{econdc}) implies that the fields $\tilde\psi_i$ are holomorphic
while the second eqn. (\ref{econdd}) is a constraint on $\tilde\psi_i$. We
shall now rescale the fields $\tilde\psi_i$ as follows
\begin{equation}
\pmatrix{\tilde\psi_1\cr\tilde\psi_2\cr\tilde\psi^3}=
\pmatrix{h\psi_1\cr\psi_2\cr h^{-1}\psi^3}\quad.
\end{equation}
$(\psi_1,\psi_2,\psi_3)$ are primary fields.
Conditions (\ref{econdc}) and (\ref{econdd}) translate to the
following conditions
\begin{eqnarray}
(\partial_{\bar z} -{\bar \mu}\partial_z + \partial_z{\bar \mu} + {\bar \rho}
(\partial_z^2 -{2\over3}\tilde{u}_2) -{1\over2} \partial_z{\bar
\rho}\partial_z +{1\over6}(\partial_z^2 {\bar \rho}))\psi_1 =0
\label{econdca}\\
(\partial_z^3 - \tilde{u}_2\partial_z - (u_3 + {1\over2}
\partial_z\tilde{u}_2))\psi_1 =0\quad,\label{econdda}
\end{eqnarray}
where ${\bar \mu}\equiv {a\over h^2}$ and ${\bar \rho}\equiv {b\over
h^4}$ are the Beltrami differentials. Further,
$$
\tilde{u}_2 = 2\partial_z \Gamma_{zz}^{~~z} + (\Gamma_{zz}^{~~z})^2 + 2a
\quad,\quad u_3 = [\partial_z - 2\Gamma_{zz}^{~~z}](\omega_z^-h)-b\quad,
$$
where we have restored dependence on $\omega_z^-$ ( even though it is
vanishing to ${\cal O}(a^2,ab,b^2)$ ). This is to explicitly
demonstrate that $u_3$ also behaves like a generalised projective
connection in the sense of \cite{GLM,BFK}.

Equations (\ref{econdca}) and (\ref{econdda}) are identical to those
obtained in \cite{GLM,BFK}.
Using arguments identical to \cite{GLM,BFK} we can show that the
these two equations are equivalent to the following OPE's.
\begin{eqnarray}
T(z)~T(w) &=& {{c/2}\over {(z-w)^4}} + {{2T(w)}\over {(z-w)^2}} +
{{\partial_wT(w)}\over {(z-w)}} +\ldots\quad,\nonumber \\
T(z)~W(w) &=& {{3W(w)}\over {(z-w)^2}} +
{{\partial_w W(w)}\over {(z-w)}} +\ldots\quad,\nonumber \\
W(z)~W(w) &=& {{c/3}\over {(z-w)^6}} + \left({{2}\over {(z-w)^4}} +
{{\partial_w}\over {(z-w)^3}} +{{(3/10)\partial_w^2}\over{(z-w)^2}}
+{{(1/15)\partial_w^3}\over{(z-w)}}\right)T(w) \nonumber \\
&&+{{16}\over{5c}}\left({2\over{(z-w)^2}}+
{\partial_w\over{(z-w)}}\right)\Lambda(w)+ \ldots
\quad,\label{eopeb}
\end{eqnarray}
where $\Lambda=T^2$. The above OPE corresponds to the semi-classical
limit ($c\rightarrow\infty$) of the OPE's of the $W_3$-algebra given
in \cite{zam} after we make the following identifications
$$
\tilde{u}_2 \longrightarrow {{24}\over c}\langle T \rangle\quad, \quad
u_3 \longrightarrow {{24}\over c}\langle W \rangle\quad.
$$

Equation (\ref{econdca}) implies the following transformation law for
a $(-1)-$ differential
\begin{equation}
\delta~\psi_1 = \xi^z \partial_z \psi_1 - (\partial_z\xi^z)\psi_1 +
\xi^{zz} (\partial_z^2 \psi_1) -
(\partial_z\xi^{zz}) (\partial_z\psi_1) -{2\over3} ((\partial_z^2 -
\tilde{u}_2 )\xi^{zz})\psi_1 \quad,
\end{equation}
where $\xi^z$ and $\xi^{zz}$ parametrise $W-$diffeomorphisms. These
parameters are holomorphic since we are in the conformal gauge.

We have demonstrated how $W-$diffeomorphisms are obtained in our
formulation. The interesting feature is that certain combinations of
the generalised connection play the role of projective connections.
So we do not need to introduce any new matter dependent fields to
describe $W$-diffeomorphisms.

\section{Conclusion and Discussions}

	In this talk, we have related the Teichmuller components of
Hitchin to the Teichmuller spaces for W-gravity. This lead to the
introduction of generalised vielbeins and connections. One can now
write out a Polyakov path integral for the W-gravity sector. Of
course, the hard part is to understand how to couple matter to
W-gravity in this formulations. The work of Schoutens {\it et
al.}\cite{schoutens} could possibly give us a hint as to how to
proceed.

	In this talk, we had restricted ourselves to the genus $g>1$
situation. What are the gauge groups for the $g=0,1$ cases? For the
bosonic case, one knows the answer ($SO(3)$ and $ISO(2)\sim
IU(1)$. This suggests the following choices: $SU(n)$ for $g=0$ and
$IU(n-1)$ for $g=1$.

In connection with $W_3$ gravity, we would like to mention the work of
Goldman who has studied the space of convex $RP^2$ projective
structures\cite{rptwo} (we shall call this space ${\cal
P}(\Sigma)$). We would like to observe that dim ${\cal P}(\Sigma)$ =
dim ${\cal T}_3(\Sigma)$. Hitchin has suggested that ${\cal
P}(\Sigma)$ is a subspace of or possibly the whole of ${\cal
T}_3(\Sigma)$. We conjecture that there exists a $RP^2$ structure in
the case of $W_3$-gravity\footnote{It has been brought to our
attention that Goldman and Choi\cite{GC} have recently shown that
${\cal P}(\Sigma) = {\cal T}_3 (\Sigma)$. We thank Pablo Ares Gastesi
for this information.}. Goldman has provided Fenchel-Nielsen
coordinates (pants decomposition) for ${\cal P}(\Sigma)$ which implies
we can discuss higher loop divergences in the context of W-strings
just as in the bosonic string\cite{gava}.

We hope that the formulation of W-gravity presented in this talk will
provide some simplifications in the description of W-gravity as well
as will lead to more insights into the geometric structures underlying
W-gravity and also provide new realisations of W-strings.\\

\noindent {\bf Acknowledgements:} I would like to thank T. Jayaraman
for a collaboration which led to the work described here.  I thank the
organisers of the Colloquium for the opportunity to present this work.
\newpage
\begin{flushleft}
{\bf References}
\end{flushleft}

\end{document}